\documentclass[twocolumn,showpacs,amsmath,amssymb,prl,graphics,graphicx]{revtex4-1}
\usepackage{bm}
\usepackage[pdftex]{graphicx}
\usepackage{dcolumn}
\begin{document}
\title{Vortices in normal part of proximity system}

\author{V.~G.~Kogan}
\email{kogan@ameslab.gov}
\affiliation{The Ames Laboratory and Department of Physics \& Astronomy,
   Iowa State University, Ames, IA 50011}

\date{\today}

  \begin{abstract}  
It is shown that the order parameter $\Delta$ induced  in the normal part of superconductor-normal-superconductor proximity system is modulated  in the magnetic field differently from  vortices in bulk superconductors. Whereas $\Delta$ turns zero at vortex centers,  the magnetic structure  of these vortices differs from that of Abrikosov's. 
  \end{abstract}

\maketitle

 The  question of superconductivity induced in the normal part (N) of   superconductor-normal-superconductor (SNS) proximity system has recently been revived by observations of vortices in N \cite{exp}.
The order parameter  $\Delta$ induced  in N    is not uniform even in zero field and is strongly suppressed nearly everywhere in  N  except the immediate vicinity of  interfaces. 
Hence, the formal problem of the order parameter distribution within vortices in N is qualitatively different from that of bulk superconductors and so do the physical properties of ``N-vortices".  These properties are of interest both for the basic physics  and for applications, enough to mention wires in superconducting magnets which are in fact SNS systems. 

Describing proximity effects,   one encounters the question of the length scale on which the induced order parameter varies. This problem is discussed in the first part of this paper for any field and temperature. In the following part, a linear combination of the eigenfunctions of an equation for $\Delta$ is constructed to represent vortices in N. In fact, the seminal work of Abrikosov on type-II superconductors suggests the form of this combination \cite{Abrikosov}. The difference, though, is that Abrikosov combined eigenfunctions of the 1st Landau level, whereas in the problem of interest here these functions are different and more general.  

As mentioned,  the induced  $\Delta$   is strongly suppressed nearly everywhere in  N  except the  vicinity of  interfaces. 
Out of this vicinity, equations of superconductivity in N can be linearized. Formally, the situation is similar to that at the upper critical field $H_{c2}$, where the magnetic field is uniform and the small $\Delta$  satisfies a linear equation 
\begin{equation}
-\xi^2\Pi^2\Delta=\Delta\,,\quad {\rm or}\quad \Pi^2\Delta=k^2\Delta\,.
    \label{linear}
\end{equation}
  at any temperature $T$ \cite{HW}.
Here, $\bm\Pi =\bm\nabla+2\pi i\bm A/\phi_0$, $\bm A$ is the vector potential, $\phi_0$ is the flux quantum, and $k^2=-1/\xi^2$.  Notwithstanding the form,  this equation differs from  the linearized Ginzburg-Landau equation (GL), in the latter the coherence length $\xi$ diverges as $T\to T_c$. At  $H_{c2}$ and $T\ne T_c$,  $\xi(T)$  is finite and is found by solving the self-consistency equation of the theory
        \begin{eqnarray}
\frac{\hbar}{2\pi T} \ln \frac{T_{c}}{T}= \sum_{\omega
>0} \left(\frac{1}{ \omega}-\frac{2\tau S}{\beta-S}  \right) ,\quad 
\beta=1+2\omega\tau\,. \qquad
\label{gap1}
\end{eqnarray}
Here,  $\hbar\omega=\pi T(2n+1)$ are Matsubara energies and  $\tau$ is the scattering time for non-magnetic impurities.  According to Helfand and Werthamer \cite{HW},
        \begin{eqnarray}
S(H_{c2})= \frac{2\beta}{ \ell q}\int_0^\infty e^{-u^2}\tan^{-1}\frac{u\ell q}{\beta } du ,\quad  
q^2=\frac{ 2\pi H}{\phi_0} , \qquad 
\label{HW}
\end{eqnarray}
where $\ell=v\tau$ is the mean-free path. 

Eq.\,(\ref{linear}) is equivalent   the Schr\"{o}dinger equation for a charge in uniform magnetic field; $H_{c2}=\phi_0/2\pi\xi^2$ corresponds to the minimum eigenvalue. The corresponding eigenfunctions belong to the first Landau level. A linear combination of these functions, constructed by Abrikosov, represents the lattice of vortices \cite{Abrikosov}.

The normal metal within the proximity system may have its own $T_{c,N}<T$ and $H_{c2,N}(T)$. We are interested here in the part of the phase diagram  outside of the region under $H_{c2,N}(T)$ (within this region the N part is superconducting and the proximity system should rather be called SS$^\prime$S). In this broad domain, the induced superconductivity  is still described by Eqs.\,(\ref{linear}) and (\ref{gap1}), however  with a more general $S(H,T,\tau)$ \cite{K85,KN}:
        \begin{eqnarray}
S(H,T,\tau)&=& \sqrt{\pi}\,{\rm Re}  \int_0^\infty ds\, \frac{(1+us^2)^\sigma }{(1-us^2)^{\sigma +1}} {\rm erfc}\,s \,,  \label{S}\\
\sigma &=&\frac{1}{2}\left(\frac{k^2}{ q^2}-1\right),\quad u=\frac{\ell^2q^2}{\beta^2}. \label{sigma}
\end{eqnarray}
Here ${\rm erfc}\,z=2\int_z^\infty e^{-z^2}dz/\sqrt{\pi}$. Solving  the self-consistency Eq.\,(\ref{gap1}) with the new $S$, one can evaluate $\xi(H,T,\tau)$ in any place of the $(H,T)$ phase diagram.

  At   $H_{c2,N}(T)$,  $\xi^2= \phi_0/2\pi H$ i.e. $k^2/q^2=-1$, and  the parameter $\sigma=-1$.  Therefore, the self-consistency equation (\ref{gap1})  in dimensionless form,
        \begin{eqnarray}
-\frac{1}{2 } \ln t= \sum_{n=0}^{\infty}\left(\frac{1}{ 2n+1}-\frac{t S}{\lambda+t(2n+1)-\lambda S}  \right) , 
\label{gap1a}
\end{eqnarray}
($\lambda=\hbar /2\pi T_{cN}\tau$ is the non-magnetic scattering parameter)
should give $H_{c2,N}(T)$ if one sets $\sigma=-1$ in $S$ of Eq.\,(\ref{S}).  Solving this numerically for the clean limit one obtains the lower curve of Fig.\,\ref{fig1}, see Appendix A.
 
 If $H\to 0$, the parameter $\sigma$ diverges, whereas $u\to 0$. It is readily shown \cite{KN} that $S$ of this case has a closed form:
        \begin{eqnarray}
S(0,T,\tau)= \frac{\beta}{ k\ell}\tanh^{-1}\frac{ k\ell}{\beta} \,. 
\label{H=0}
\end{eqnarray}
Solving  numerically Eq.\,(\ref{gap1}) with $S(0,T,\tau)$ one obtains the decay length $\xi=1/k$ of the  order parameter in the normal part of proximity systems \cite{DeGennes,K82} in zero field.
   \begin{figure}[htb]
   \includegraphics[width=7.5cm]{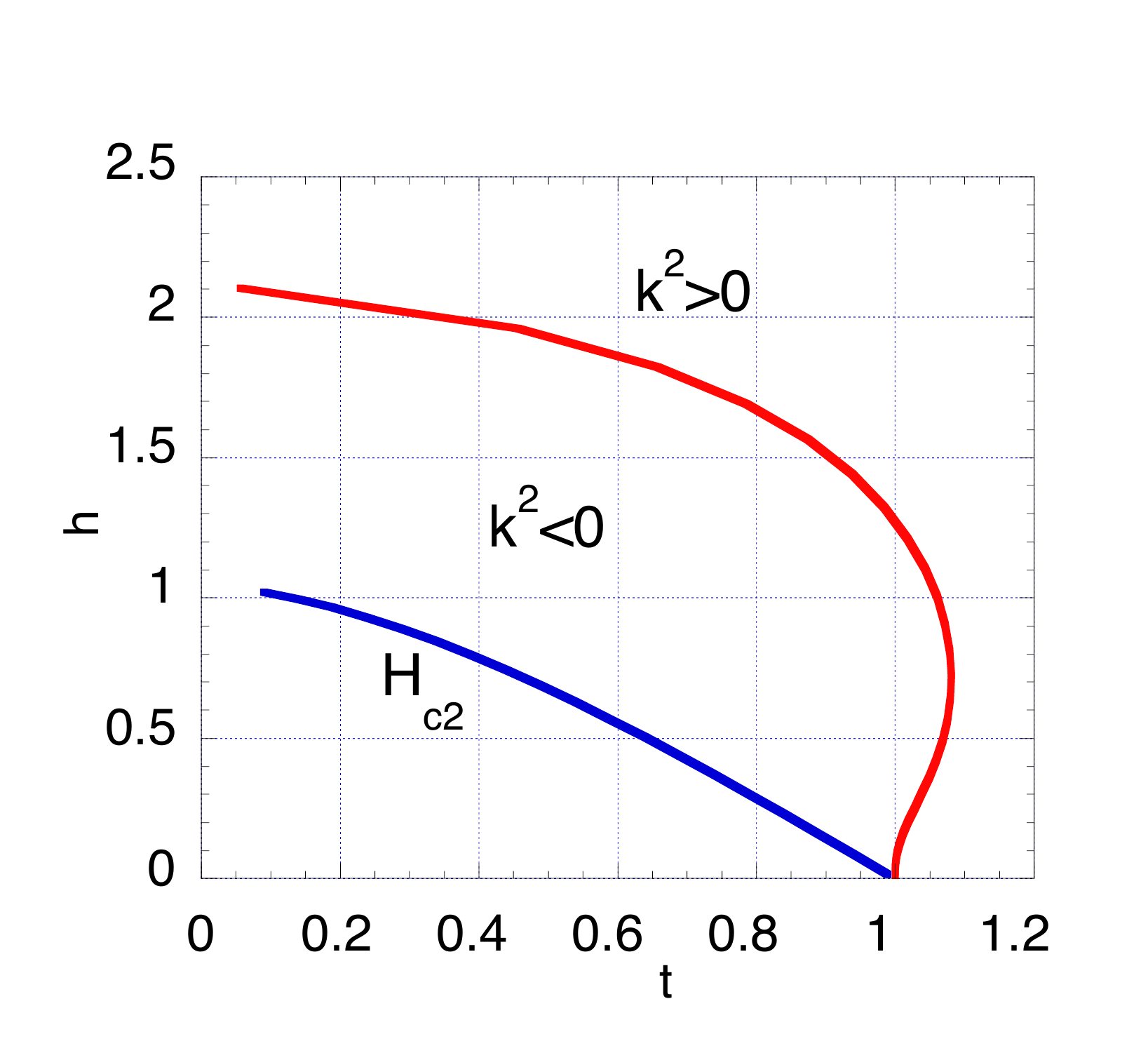}
 \caption{(Color online) The lower curve is  $H_{c2}$ of the clean limit in units $ 2\pi T_{cN}^2\phi_0/\hbar^2v^2$; at this curve $ k^2=-q^2$ and $\sigma=-1$.  At the upper  curve $k^2=0$ and $\sigma=-1/2$.   Between the upper and lower curves, $k^2$ is negative. Above and to the right of the upper curve, $k^2$ is positive. 
 }
 \label{fig1}
 \end{figure}
   
Thus,   $k^2=-\xi^2<0$ at the curve $H_{c2}$ whereas it must be positive in zero field at $t>1$, where it describes $\Delta$ attenuation in the N phase. This suggests that a curve   exists on the plane $(H,T)$  where $k^2=0$ \cite{K85}. This question is addressed  by setting $\sigma=-1/2$ in 
 $S$ of Eq.\,(\ref{S}) and solving the latter for $q^2(t)$.  This curve evaluated numerically   for the clean limit  is the upper one in Fig.\,\ref{fig1}. 
 
 The question then arises about  behavior of the normal metal in the part of the $(H,T)$ plane where $k^2$ is negative (between the curves of Fig.\,\ref{fig1}) and out of it where $k^2>0$. To address this we look at  eigenfunctions $\Delta(x,y)$ of the equation $\Pi^2\Delta=k^2\Delta$. 
Choosing  $\bm A=-Hy\,{\hat {\bm x}}$ we have
        \begin{eqnarray}
 \left(\frac{\partial}{\partial x}+i\frac{2\pi H}{\phi_0} y\right)^2\Delta +\frac{\partial^2\Delta}{\partial y^2} =k^2\Delta\,.
\label{Schrod}
\end{eqnarray}
The equation does not contain $x$ explicitly, so that
        \begin{eqnarray}
\Delta=\Delta_0 e^{i p x}\chi(y)\, 
\label{form}
\end{eqnarray}
with $\chi(y)$ satisfying
        \begin{eqnarray}
d^2\chi/dy^2 - q^4(y+p/q^2)^2\chi =k^2 \chi\,. 
\label{form_a}
\end{eqnarray}
 In terms of    $\tilde y=y+p/q^2$  the general  solution is:
     \begin{figure}[htb]
   \includegraphics[width=8cm]{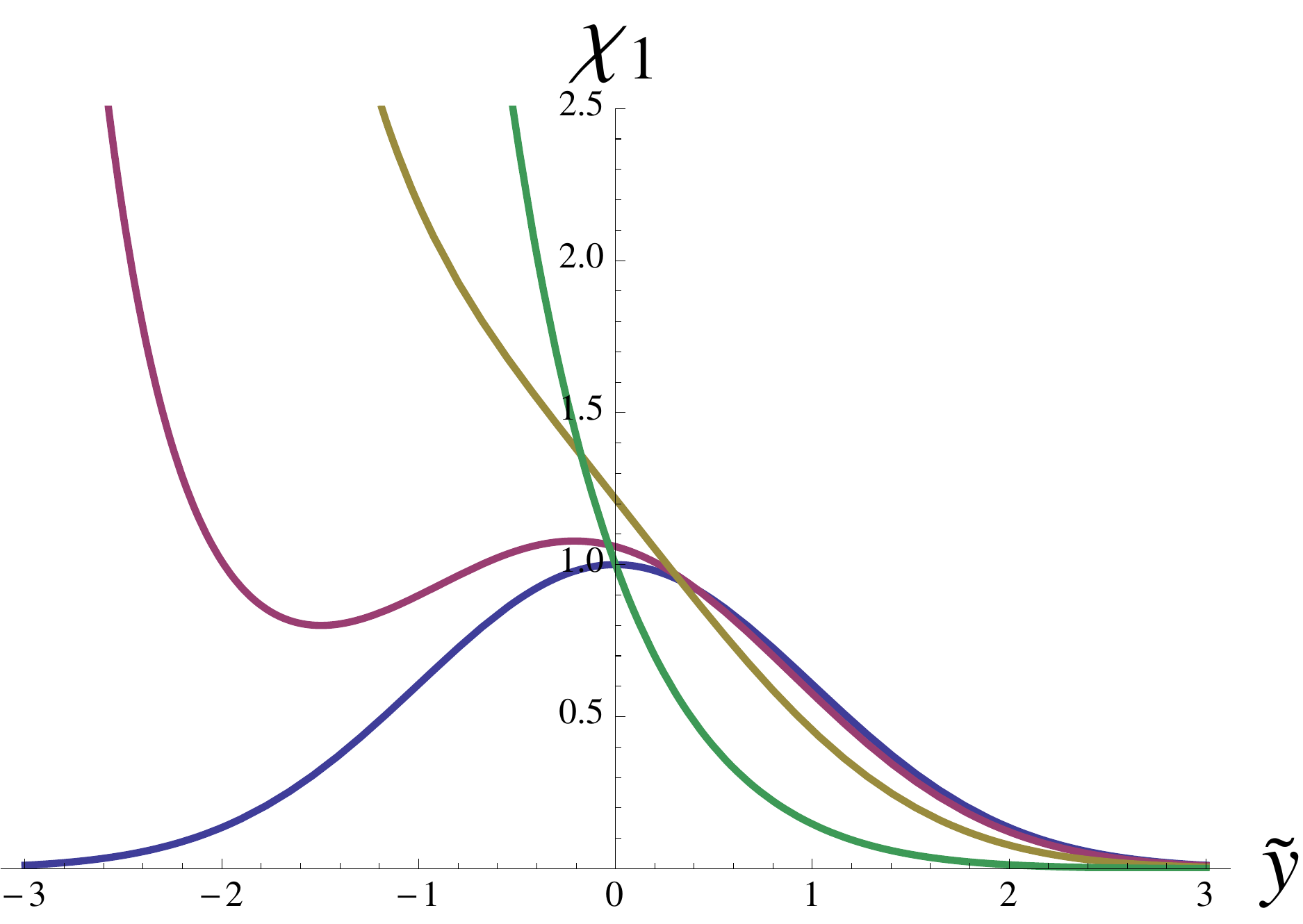}
 \caption{(Color online)  $\chi_1(\sigma,\tilde y)$ for $q=1$ in units $2\pi T_c/\hbar v$.  Ordering curves by  their left edges clockwise: $\sigma=-1$ ($H_{c2}$), $-0.9$ (with a minimum at the left), $-0.5$ ($k=0$), and 1.     }
 \label{fig2}
 \end{figure}

         \begin{eqnarray}
\chi& =& C_1\chi_1+C_2\chi_2\,,\nonumber\\
\chi_1&=&2^{(\sigma+1)/2} e^{-q^2\tilde y^2/2}\,{\cal H}\left(-\sigma-1,\,\,q\tilde y\right)\,,\nonumber\\
\chi_2&=&2^{-\sigma/2} e^{q^2\tilde y^2/2}\,{\cal H}\left(\sigma,\,\, iq\tilde y\right)\,, 
\label{general}
\end{eqnarray}
with arbitrary constants $C_{1,2}$  and    $\sigma$   of Eq.\,(\ref{sigma}). The Hermite  functions ${\cal H}(\sigma,\,\, w )$ can be expressed in terms of the parabolic cylinder functions  and reduce to Hermite polynomials for $\sigma=0,1,2,\,... $ \cite{Abramowitz}.  

Note that $\chi_1 $
with $\sigma$ being a negative integer are the harmonic-oscillator wave-functions  that go to 0 as $\tilde y\to\pm\infty$; these are the eigenfunctions of the  Landau levels. We are interested here in the part of the phase diagram where $\sigma > -1$ where  $\chi_1 (\tilde y)$ is real, diverges as $\tilde y\to -\infty$, and goes to 0 as $\tilde y\to +\infty$, see Fig.\,\ref{fig2}. For  symmetric SNS systems,  $\chi_1 $ should be discarded. 

\begin{figure}[h]
   \includegraphics[width=8cm]{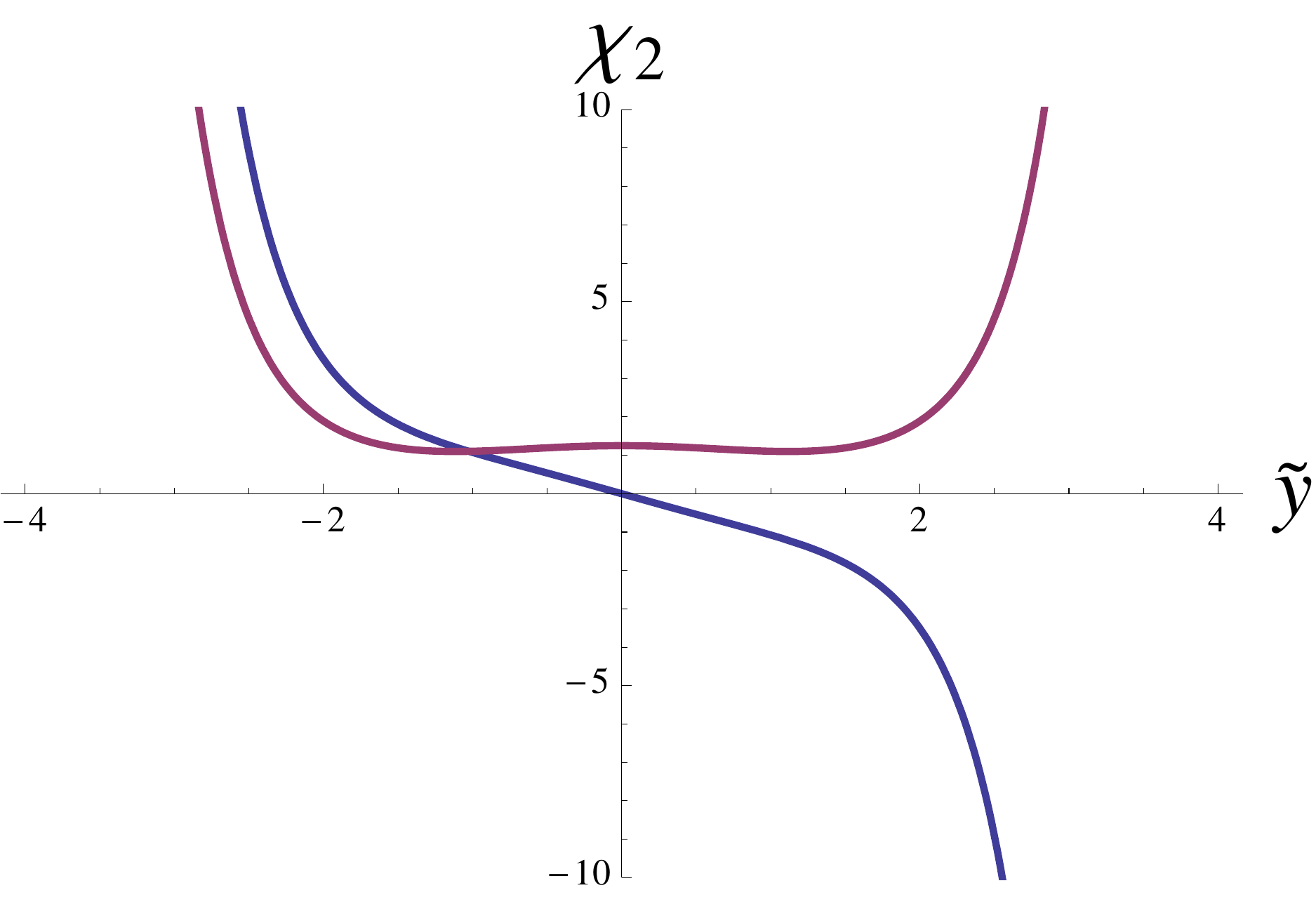}
 \caption{(Color online) The function $\chi_2(\sigma,\tilde y)$ for $\sigma=-0.7$ and $q=1$. Re$[\chi_2(\tilde y)]$ is even,  Im$[\chi_2(\tilde y)]$ is odd function of $ \tilde y $.   }
 \label{fig3}
 \end{figure}

On the other hand,  $\chi_2$ in general has both real and imaginary parts. An example of $\chi_2(-0.7,\,\,1.,\,\, \tilde y)$ is shown in Fig.\,\ref{fig3}. Both real and imaginary parts diverge at large $\tilde y$. 
  $\chi_2$ should  be taken into account in finite samples, unlike  the infinite ones where it should be disregarded. 

Consider now a normal metal layer between two thick superconducting banks forming the SNS proximity  junction. The N slab of a thickness $W$ is infinite in $x$ and $z$ directions whereas $-W/2<y<W/2$. 
The temperature of the system $T_{cN}< T<T_{cS}$. In zero field Eq.\,(\ref{Schrod}) describes   exponential decay of  the order parameter away of interfaces, $\Delta\propto \cosh y/\xi$. In the 
 magnetic field along $z$ which is less than $H_{c1,S}$, the field is confined to the N domain, whereas the S banks are in the Meissner state (assuming the London penetration depth    $\lambda_S\ll W$). Since superconductivity is induced in  N  by proximity with $S$, one  expects vortices to be nucleated within the N layer. In small enough fields, vortices should form a periodic chain in the slab middle  at $y=0$.

The N slab is uniform in the $x$ direction, so that the parameter $p$ in the solution (\ref{form}) can take any value. Consider a linear combination
        \begin{eqnarray}
\Delta= e^{i p x}\chi_2(y+p/q^2) + e^{-i p x}\chi_2(y-p/q^2)\,,
\label{combination}
\end{eqnarray}
where the overall constant $\Delta_0$ is omitted. It is clear that if $\Delta(x_0,0)=0$, the zero should be repeated with the period $\delta x_0 = \pi/p$. 
 If the penetration depth into the $S$ banks is small relative to  $W$, the flux quantization would have given $\delta x_0W H =\phi_0$ and the parameter $p$ would be
        \begin{eqnarray}
p= \frac{\pi W H  }{\phi_0}= \frac{q^2W}{2}\,.
\label{quantization}
\end{eqnarray}
Unlike the problem of $H_{c2,N}(T)$ where
  $\xi_0= \hbar v/2\pi T_{cN}$ was adopted as a natural unit length, it is   convenient here to normalize lengths to $W/2$, the half-width of the N layer. Then, the dimensionless $p=q^2 $. Since the RHS of Eq.\,(\ref{combination}) is dimensionless, we keep the same notation   $x,y,p,q$ as for their dimensional counterparts. 

\begin{figure}[h]
   \includegraphics[width=8cm]{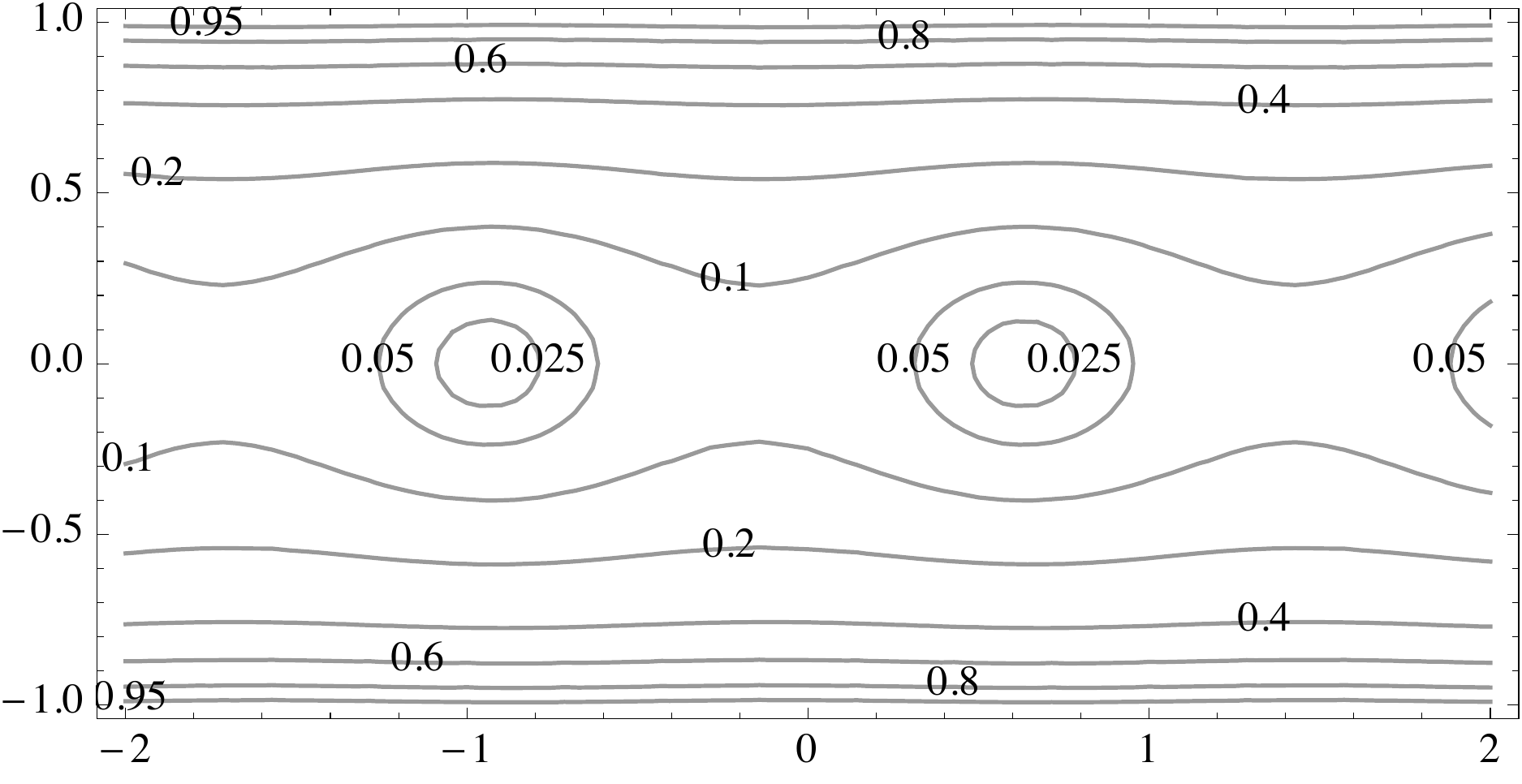}
 \caption{(Color online) Lines of the constant  order parameter modulus  according to Eq.\,(\ref{combination}) with $\sigma=0.2$, $q=\sqrt{2}$, $p=2$ for $-2<x<2$ and $-1<y<1$.   }
 \label{fig4}
 \end{figure}

\begin{figure}[h]
   \includegraphics[width=7.cm]{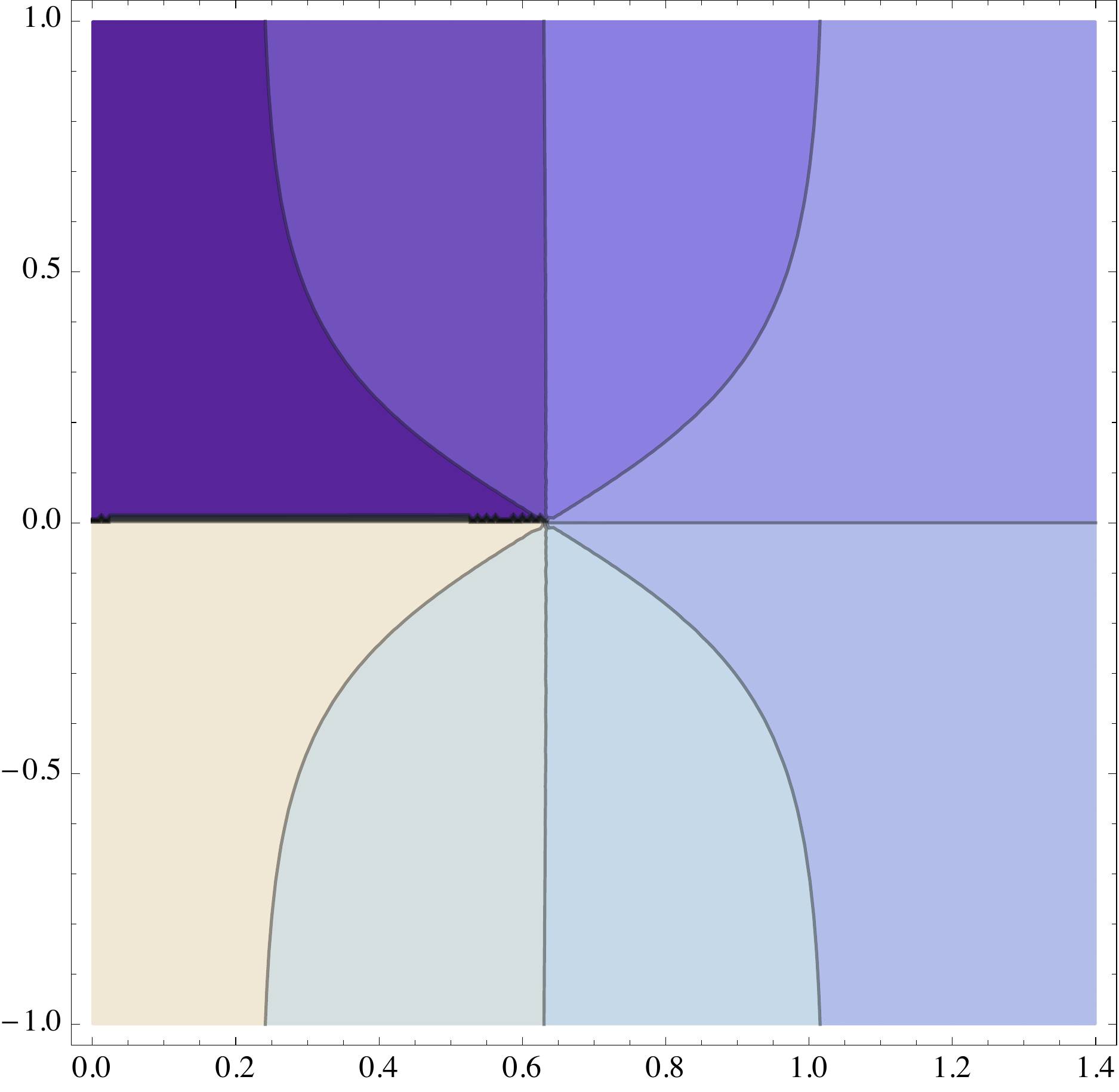}
 \caption{(Color online) Contours of the constant phase  with the step $\pi/4$ in  clock-wise order for   the vortex at $x\approx 0.6$ and $y=0$ of Fig.\,\ref{fig4}. The phase jumps by $2\pi$ at the straight ray from the vortex center to the left.  }
 \label{phase}
 \end{figure}

The structure of the solution (\ref{combination}) is illustrated in Fig.\,\ref{fig4}  where the modulus $|\Delta(x,y)|$ is plotted for  $W=2$, $\sigma=0.2$, $q^2= p=2$. As predicted, the distance between 
singularities (vortices) is $\delta x_0 = \pi/p\approx 1.57$. This means that the flux quantization holds indeed for vortices in the $N$ layer, which is not self-evident in advance.  
 The solution shown is normalized as to have $\Delta\approx 1$ at the interfaces $y=\pm 1$, i.e.,  $\Delta_0 $ is  set equal to inverse of the RHS of Eq.\,(\ref{combination}) taken at $y=1$. 
The phase of this solution near one of the vortices is shown in Fig.\,\ref{phase}. 

\begin{figure}[h]
   \includegraphics[width=7.5cm]{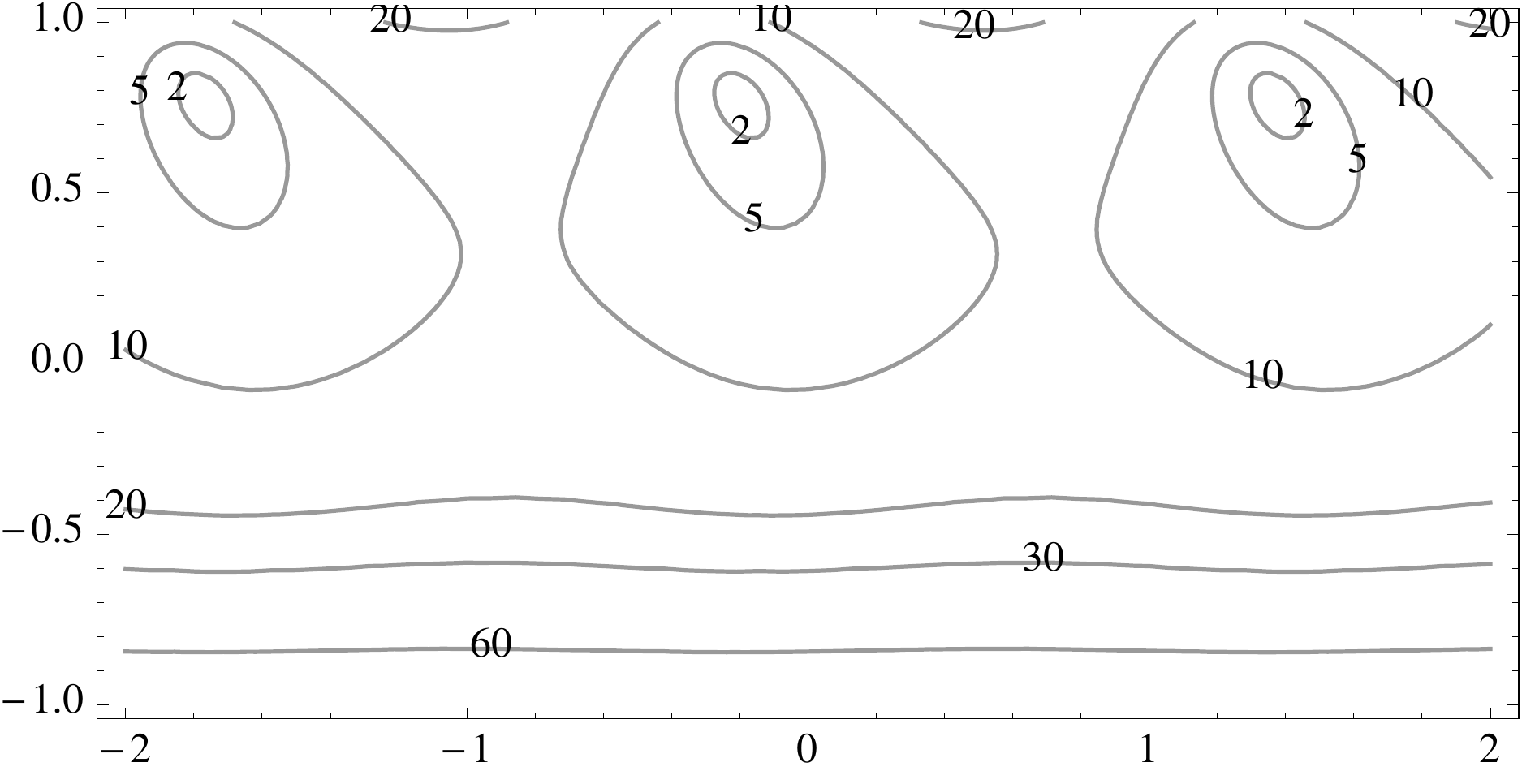}
 \caption{(Color online) Contours of $|\Delta |=$  constant  for the order parameter of Eq.\,(\ref{combination1}) with $\sigma=-0.8$ and $p=q^2=2$.   }
 \label{example}
 \end{figure}

It should be noted that the form (\ref{combination}) of the order parameter is not the only  possibility. One can consider various linear combinations with different complex coefficients which all satisfy   $\Pi^2\Delta=k^2\Delta$. The choice of these coefficients is dictated by the boundary conditions  (the form of the S banks and the distribution of the order parameter on these banks). An example of  $|\Delta |$ for 
        \begin{eqnarray}
\Delta= e^{i p x}\chi_2(y+p/q^2) + 5(1+i)e^{-i p x}\chi_2(y-p/q^2)\qquad
\label{combination1}
\end{eqnarray}
is shown in Fig.\,\ref{example}.  
In this example $\Delta$ is not normalized and it  clearly does not satisfy the boundary conditions $\Delta_N(y=\pm 1)=\Delta_S$, but it shows qualitatively that this type of linear combinations might be useful in describing the proximity effect in asymmetric SNS$^\prime$ systems with different S-banks  for which vortices in $N$ tend to be closer to the bank with smaller order parameter.  The value $\sigma=-0.8$ is chosen to illustrate that in the domain of negative $k^2$ (between two curves of Fig.\,\ref{fig1}) vortices (better to say regions with closed current lines) occupy larger areas at the same field than in the case $k^2>0$ of Fig.\,\ref{fig4}.

Hence, vortices appear at the bottom of the  suppressed order parameter valley. They have  normal cores in a sense that   $\Delta=0$ at the  center of each  vortex and the phase changes by $2\pi$ if one circles the   center. 
Still, they differ from their   Abrikosov ``brethren".  The order parameter changes differently with the distance from the center along $x$ or $y$ directions. Unlike Abrikosov's case, one cannot define the core size as the distance from the center to a place with depairing current density.  Besides, the self-energy of these vortices should be quite small because they appear in the region where the order parameter is strongly suppressed even in zero field. 
The magnetic field is practically constant in the N layer (it is exactly constant within our model).  In particular, this implies that methods of observing vortices  based on detecting the vortex field (decoration or scanning SQIUD microscopy) will probably not work. On the other hand, STM that  probes the order parameter value should discern zeros of $\Delta$. In fact, the recent STM data show vortices between superconducting Pb islands separated by the normal wetting layer \cite{exp}. 

There are many questions  remain on properties of vortices within   domains of proximity induced   superconductivity. Currents through the SNS sandwich in magnetic field should cause vortex motion. Is this motion  overdamped and if it is, what is the drag coefficient? The Bardeen-Stephen  formula is unlikely to work since it is not even clear what plays the role of the vortex core size  in the normal metal. 

An interesting question concerns superconducting fluctuations in the N phase.   According to pioneering results of  Schmid \cite{Schmid} and Prange \cite{Prange} based on linearized GL equation, the diamagnetic susceptibility $\chi_d$ in the normal phase is proportional to $\xi$. In particular, in zero field, $\chi_d$ diverges as $T$ approaches $T_c$ from above. Here, a method to evaluate $\xi(H,T)$ is offered for any place in the $(H,T)$ phase diagram. It would be of interest to look at possible differences in diamagnetic susceptibility within the region where $k^2=-1/\xi^2$ is negative (between the curves of Fig.\,\ref{fig1}) and out of it where $k^2$ is positive. \\

 The author is grateful to V. Dobrovitski, L. Bulaevskii, S. Bud'ko,  R.~Prozorov,  P.~Canfield, D.~Finnemore, M.~ Hupalo  for helpful discussions. The Ames Laboratory is supported by the Department of Energy, Office of  Basic Energy Sciences, Division of Materials Sciences and Engineering under Contract No. DE-AC02-07CH11358.\\

{\it {\bf Appendix A.}} For the numerical work the integral (\ref{S1}) is rewritten to account for the branch point at $s=1/\sqrt{u}$:
        \begin{eqnarray}
S=\sqrt{\frac{\pi} { u}} \int_0^1 d\eta \frac{(1+\eta^2)^\sigma}{(1-\eta^2)^{\sigma+1}}  
  \left[{\rm erfc}\frac{\eta}{\sqrt{u} }-\cos(\pi\sigma)\,{\rm erfc}\frac{1}{\eta\sqrt{u}}\right]. \nonumber\\
\label{S1}
\end{eqnarray} 
For the calculation of $H_{c2,N}(T)$, it is convenient to measure length in units of $\hbar v/2\pi T_{cN}$. Then, we have:
        \begin{eqnarray}
\sqrt{u}=\frac{ q }{\lambda+t(2n+1) }\,,\quad   q^2=\frac{\hbar^2v^2  H}{2\pi T_{cN}^2\phi_0} \approx \frac{H}{H_{c2,N}(0)}  ,\qquad
\label{mu,q a}
\end{eqnarray}
where $\lambda=\hbar /2\pi T_{cN}\tau$ is the non-magnetic scattering parameter, 
$H_{c2,N}(0)$ is the zero-$T$ clean limit upper critical field, and $t=T/T_{cN}$. \\

{\it {\bf Appendix B.}} The assumption of a finite $T_{cN}$ in the main text  is in fact not necessary. However, the formal treatment of the case $T_{cN}=0$ should take into account that   $\Delta=0$ when the effective coupling is zero. Nevertheless, proximity with S results in non-zero Green's functions $F(\omega)$ in the normal metal. This leads to  $\beta-S=0$ and to different exponential decay lengths of $F$ for different $\omega=\pi T(2n+1)/\hbar$. The longest length corresponds to $n=0$, i.e. to $\omega=\pi T$, so that calculating the depth of   pairs penetration one can disregard all $n\ne 0$ \cite{Kupr,K85}.  
   \begin{figure}[h]
   \includegraphics[width=7.cm]{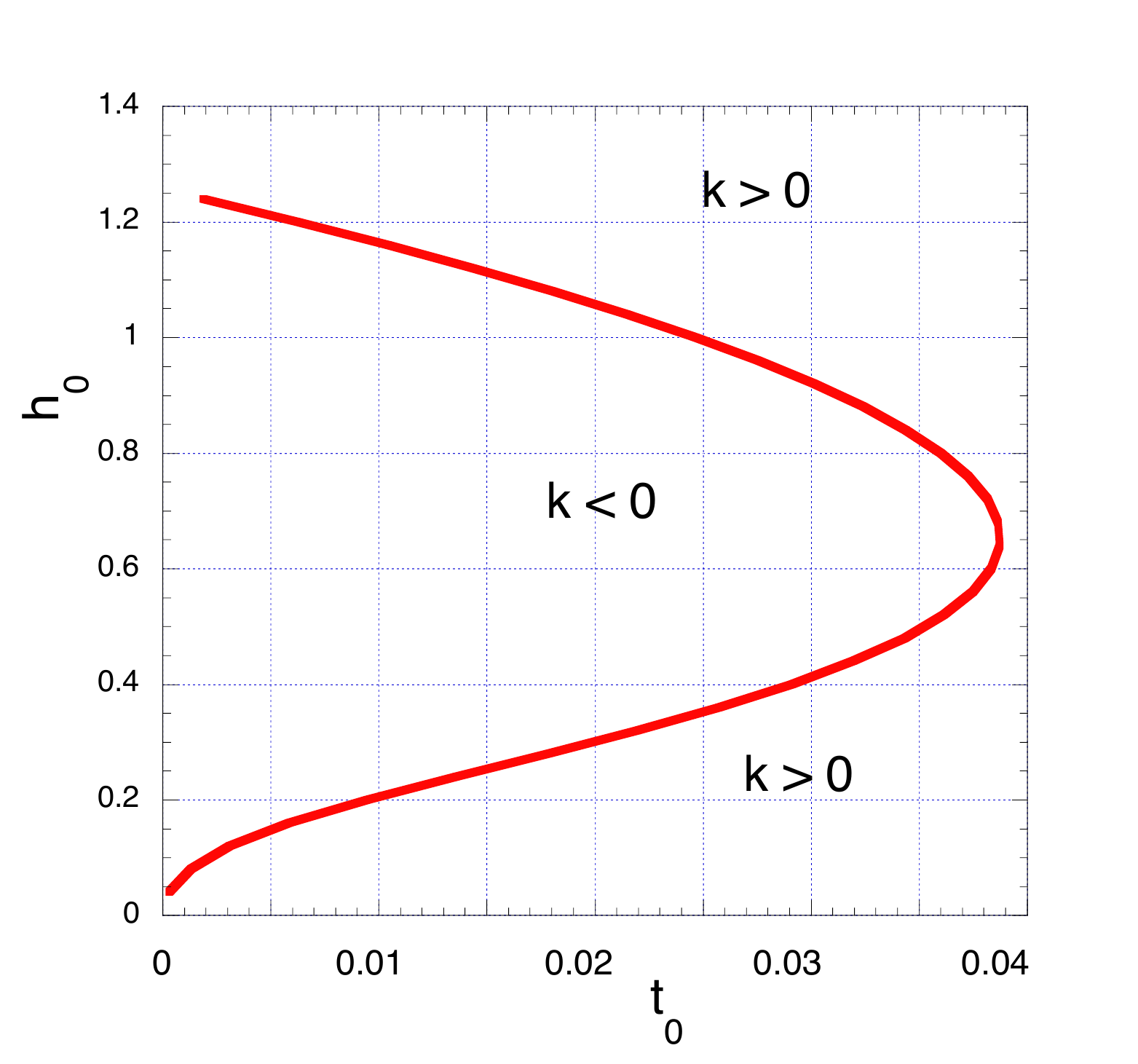}
 \caption{(Color online) The curve $k=0$ for $T_{cN}=0$. Note that this curve holds for any mean-free path; the actual temperature and field are $T=\hbar t_0/2\pi\tau$ and $H=\phi_0h_0/2\pi\ell^2$. Hence at $(H,T)$ plane, on approaching  the clean limit, this curve shrinks to the origin  so that $k>0$ everywhere. On the other hand, the domain of   $k<0$ expands with increasing scattering. 
  }
 \label{fig7}
 \end{figure}
 
Since in this situation there is no standard energy scale related to  $\Delta $ or $T_c$ (and no length scale  $\hbar v/T_c$), one can use the 
following  reduced temperature and field:
        \begin{eqnarray}
t_0=\frac{{2\pi\tau}}{\hbar}T , \quad  h_0=\frac{2\pi\ell^2}{\phi_0}H.  \label{t0}
\end{eqnarray}
In these variables, $\beta=1+t_0$ and $u=h_0/(1+t_0)^2$. To find $k(t_0,h_0)$ one has to solve $1+t_0=S(u,\sigma)$ with $S$ taken at $n=0$. Consider, as an example,  the curve $k=0$ at which $\sigma=-1/2$. Using the form (\ref{S1}), we have
        \begin{eqnarray}
S_0= \sqrt{\frac{\pi}{ u}} \int_0^1  \frac{d\eta}{\sqrt{1-\eta^4}}  \,
 {\rm erfc}\frac{\eta}{\sqrt{u}}\,. 
\label{S0}
\end{eqnarray}
This integral is expressed in terms of generalized hypergeometric functions, which are easily treated with the help of Mathematica.  Solving numerically $1+t_0=S_0(u,\sigma)$ one obtains the curve of Fig.\,\ref{fig7}.

  \end{document}